\documentclass[twocolumn,showpacs,amsmath,amssymb]{revtex4}

\usepackage{graphicx}
\usepackage{dcolumn}
\usepackage{bm}
\newcommand{\etal}{\emph{et al.}}
\newcommand{\be}{\begin{equation}}
\newcommand{\ee}{\end{equation}}
\newcommand{\bfig}{\begin{figure}}
\newcommand{\efig}{\end{figure}}

\begin{document}
\title{Unusual Nernst effect suggestive of time-reversal violation in the 
striped cuprate La$_{2-x}$Ba$_x$CuO$_4$
}
\author{Lu Li$^{1*}$, N. Alidoust$^1$, J. M. Tranquada$^2$, G. D. Gu$^2$, and N. P. Ong$^1$
}
\affiliation{
$^1$Department of Physics, Princeton University, Princeton, NJ 08544\\
$^2$Brookhaven National Laboratories, Upton, NY 11973
}

\date{\today}
\begin{abstract}
The striped cuprate La$_{2-x}$Ba$_x$CuO$_4$ ($x=\frac18)$ undergoes several
transitions below the charge-ordering temperature $T_{co}$ = 54 K. From 
Nernst experiments, we find that, below $T_{co}$, there exists a large,
anomalous Nernst signal $e_{N,even}(H,T)$ that is symmetric in field $H$, 
and remains finite as $H\to 0$. The time-reversal violating signal
suggests that, below $T_{co}$, vortices of one sign are spontaneously created to relieve 
interlayer phase frustration. 
\end{abstract}


\maketitle                   
In the cuprates, there is increasing evidence that time-reversal 
invariance (TRI) is broken over a large portion of the phase diagram. 
Following a prediction in cuprates~\cite{Varma}, signatures of TRI-violation
were obtained in angle-resolved
photoemission~\cite{Campuzano} and polarized neutron scattering experiments~\cite{Bourges}. 
Recently, polar Kerr rotation measurements~\cite{Kapitulnik,Shen} and 
polarized neutron scattering experiments~\cite{Greven} have uncovered firmer
evidence for TRI-violating states in several cuprates.

The cuprate La$_{2-x}$Ba$_x$CuO$_4$ at doping $x\simeq\frac18$ undergoes a 
remarkable series of electronic phase transitions starting at the charge-ordering
temperature $T_{co}$ (54 K) and followed by the spin-ordering temperature
$T_{so}$ (40 K) and the Berenzinski-Kosterlitz-Thouless (BKT) transition $T_{BKT}$ (16 K)
~\cite{QLi07,Hucker08,TranquadaPRB08,Hucker11}. 
Below 5 K, 3D superconductivity is established. We have observed
an unusual zero-field Nernst effect signal that appears 
below $T_{co}$.
In principle, such a zero-field Nernst signal is
forbidden in a material that has TRI.  
We discuss the implications of its appearance below the charge ordering temperature $T_{co}$.

\begin{figure}[h]
\includegraphics[width=8 cm]{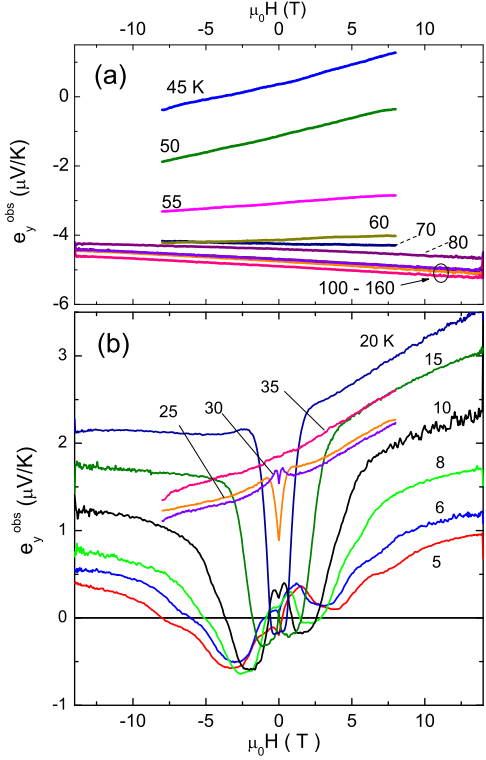}
\caption{\label{EyH} (color online)
Traces of the observed (raw) Nernst signal $e^{obs}_y(H,T)$ vs. applied field $H$ 
at selected $T$ from 160 to 45 K (Panel a) and below 35 K (b).
The curves in Panel (a) are nominally linear in $H$, with an intercept
at $H=0$ that comes from ``pick up'' of the longitudinal thermopower $S$ due to
contact misalignment. The Nernst coefficient $\nu$ is obtained from the slope
near $H=0$. Below 30 K (Panel b), the curves of $e^{obs}_y(H,T)$ 
display prominent oscillatory features at low $H$ which we identify with an
anomalous field-symmetric Nernst signal $e_{N,even}(H,T)$.
}
\end{figure}



Nernst effect measurements were carried out on La$_{2-x}$Ba$_x$CuO$_4$ crystals 
with $x = \frac18$ (LBCO-$\frac18$). We cut crystals (2, 0.7, 0.2 mm$^3$ along 
the crystal axes $\bf a,b,c$, respectively) 
from a boule and polished the faces until the normal to the broadest face was aligned with 
$\bf c$ to within $\pm 0.5^{\rm{o}}$. For each curve of the Nernst signal
vs. the applied field $\bf H$, we made dual measurements at two temperature gradients ($-\nabla T$ = 0.5 K/mm and 0.7 K/mm)
to check for linearity and reproducibility. The field was swept slowly at rates 0.2 T/min to 0.5 T/min.
The measured thermal conductivity $\kappa$ has a relatively weak $T$ dependence between 10 and 60 K
(varying between 6 and 7.2 W/Km).
In our geometry, $-\nabla T$ is applied $||\mathbf{a}$ in 
the LTT phase (with axes $\bf\hat{x}||a$, $\bf\hat{z}||c$). With $\bf H||\hat{z}$, the
voltage $V_y$ observed along $\bf \hat{y}||b$ gives the observed Nernst signal, 
$e_y^{obs}(H,T) \equiv V_y(H,T)/(|\nabla T|d)$ with
$d$ the voltage-contact spacing (we use little ``$e$'' to denote the Nernst 
electric-field $E_y$ divided by $|\nabla T|$). In Nernst experiments, $e_y^{obs}$ is often contaminated by
unavoidable pickup of the longitudinal thermopower signal caused by slight lead misalignment. 
We show that the anomalous signal is distinct from this pickup.

In Fig. \ref{EyH}, we show the observed Nernst signal at selected $T$ from 160 K to 45 K
(Panel a) and for $T \le$35 K (Panel b). Above 35 K, $E_y$ is nominally linear in $H$
with a zero-field intercept that we identify with the zero-$H$ thermopower $S(0)$.
The tilt of the curves is the conventional field-antisymmetric Nernst signal.
Below the charge ordering at $T_{co}$ = 54 K, however, $e_y(H,T)$ displays 
anomalous features which become prominent below 30 K
(Panel b). The sharp, zero-field anomaly visible at 30 K grows steeply 
in the negative direction (relative to the
zero-$H$ value at 35 K) as $T$ falls to 25 K. At 20 K, the anomaly assumes
the shape of a narrow $H$-symmetric trench of full-width $\sim$2 T. As $T$ decreases from
20 to 6 K, the trench width broadens rapidly to 15 T. At low $T$, we observe
new structures appearing at lower fields.

\begin{figure}[t]
\includegraphics[width=8 cm]{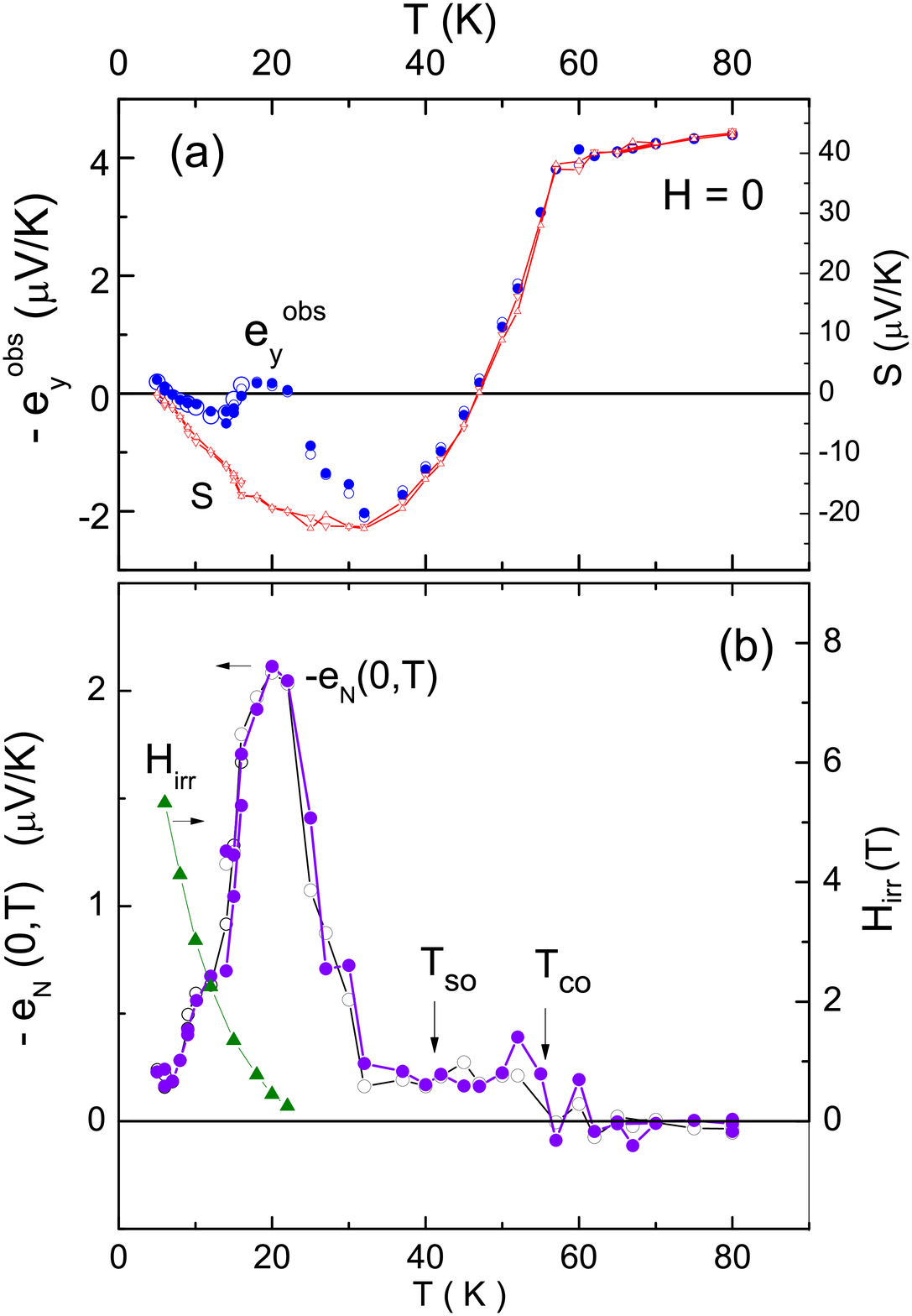}
\caption{\label{EyT_ST} (color online)
Subtraction of the thermopower to extract the anomalous Nernst signal. Panel (a) compares
the observed Nernst signal $E^{obs}_y(0,T)$ with the thermopower $S(0,T)$ at $H$=0. 
The Nernst results include two sets of data taken with $|\nabla T|\sim$ 2 K/cm (solid symbols) 
and $\sim$ 4 K/cm (open). 
By fixing the scaling number $k$= -9.8, the two curves can be superposed in the interval
50-90 K. Below 32 K, the curves strongly deviate from each other. The difference is
identified with the zero-$H$ anomalous Nernst signal $E_N(0,T)$, which is plotted in Panel (b). 
For 30$<T<$ 50 K, $|E_N(0,T)|$ is small (0.2 $\mu$V/K) but well-resolved.
Below 30 K, it rises abruptly to a prominent peak at 20 K before decreasing to zero near 5 K.
The irreversibility field $H_{irr}$ measured by torque magnetometry is plotted as 
solid triangles.
}
\end{figure}

Generally, the Nernst electric field $E_y$ is antisymmetric in $H$, vanishing at $H$ = 0.  
Initially, we attributed the zero-$H$ signal in Fig. \ref{EyH} to pickup
of the longitudinal signal $S$. This assumption is valid above 60 K. However, below
54 K, a distinct field-even signal distinct from $S(H,T)$ becomes resolvable.
To show this, we have measured the thermopower $S(H,T)$ 
simultaneously with the Nernst signal.
Figure \ref{EyT_ST}(a) displays the $T$ dependence of $e^{obs}_y$ and $S$ 
measured in zero field. We find that $S$ is positive 
above $T_{co} \sim$ 54 K, decreases rapidly below 54 K, becoming 
negative below 45 K. At lower $T$, $S$ attains a broad minimum at 30 K 
before vanishing near $T_c$ = 5 K.

First, we compare the zero-$H$ values of the observed Nernst
signal $e^{obs}_y(0,T)$ (circles in Fig. \ref{EyT_ST}a) and $S(0,T)$ (solid curve)
over a broad interval of $T$. 
Above 54 K, the two quantities track closely. Multiplying the former by
a scaling number $k$, we may superpose the two curves (Fig. \ref{EyT_ST}a). 
The value of $k$ (-9.8) implies that the voltage contacts were slightly misaligned  
by $\sim$130 $\mu$m along $\hat{\bf{x}}$. Below $T_{co}$, the two quantities
deviate significantly. In contrast to the curve of $S$,
$e^{obs}_y(0,T)$ oscillates vs. $T$, changing sign four times. 
With $k$ = -9.8, we may isolate intrinsic Nernst signal 
$e_N(H,T)$ at finite $H$ by substracting off the 
thermopower signal, viz.
\begin{equation}
e_N(H,T) = e^{obs}_y(H,T) - kS(H,T).
\label{eq:eN}
\end{equation}

The quantity $e_N(0,T)$ in zero $H$, plotted in Fig. \ref{EyT_ST}b, is of main interest.
In the interval 30-54 K, the
magnitude of $|e_N(0,T)|$ equals 0.2 $\mu$V/K, which is easily
resolved in our experiment. Below 30 K, it rises steeply 
to a prominent maximum of 2.2 $\mu$V/K at 20 K before falling 
to zero near 5 K. The prominent peak, which is very sensitive to $H$, 
is the cause of the trench feature bracketing
$H=0$ in the curves of $e^{obs}_y$ vs. $H$ plotted in Fig. \ref{EyH}.

\begin{figure}[t]
\includegraphics[width=7.5 cm]{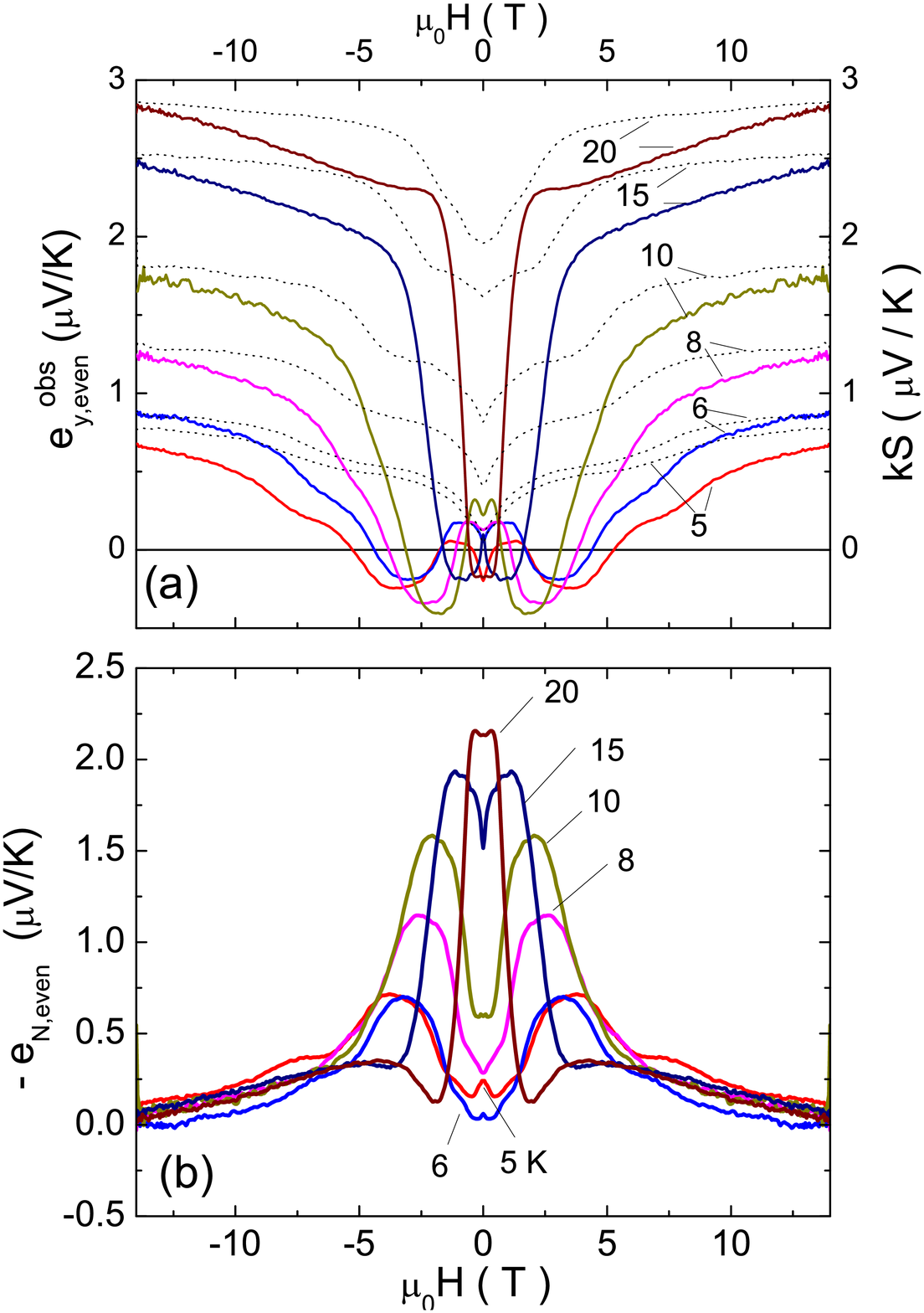}
\caption{\label{EyHSH} (color online)
Panel (a): Comparison of the raw, field-symmetrized, Nernst signal $e^{obs}_{y,even}(H,T)$
(solid curves) with the thermopower $S(H,T)$ (scaled by $k$ = -9.8, dashed curves) at selected $T\le$ 20 K. 
Note that $S(H,T)$ is actually negative below 40 K (at all $H$ shown).
The two sets of curves have very different field dependences. Panel (b) displays the curves of the
intrinsic field-symmetrized Nernst signal $e_{N,even}(H,T)$ obtained by
subtracting the two sets of curves (see Eq. \ref{eq:eN}). The oscillatory features
are absent in $S(H,T)$. At large $H$, $e_{N,even}(H,T)$ is suppressed to zero.
}
\end{figure}



It is also instructive to examine the field-symmetrized 
form of the observed Nernst signal 
$e^{obs}_{y,even}(H)= \frac12[e^{obs}_y(H) + e^{obs}_y(-H)]$ which admixes
$e_{N,even}$ and $S$. At 20 K, $e^{obs}_{y,even}(H,T)$
displays a deep trench centered at $H$=0 (Fig. \ref{EyHSH}a). As $T$ decreases to 5 K, the trench
broadens rapidly. For comparison, we also plot the curves of $S(H,T)$
(scaled by the parameter $k$). The features
in the field profiles are clearly distinct 
in the two sets of curves. This difference provides strong evidence 
that the Nernst signal $e_N(H,T)$ has an intrinsic
field-even component that is distinct from $S(H,T)$.

\begin{figure}[t]
\includegraphics[width=7.5 cm]{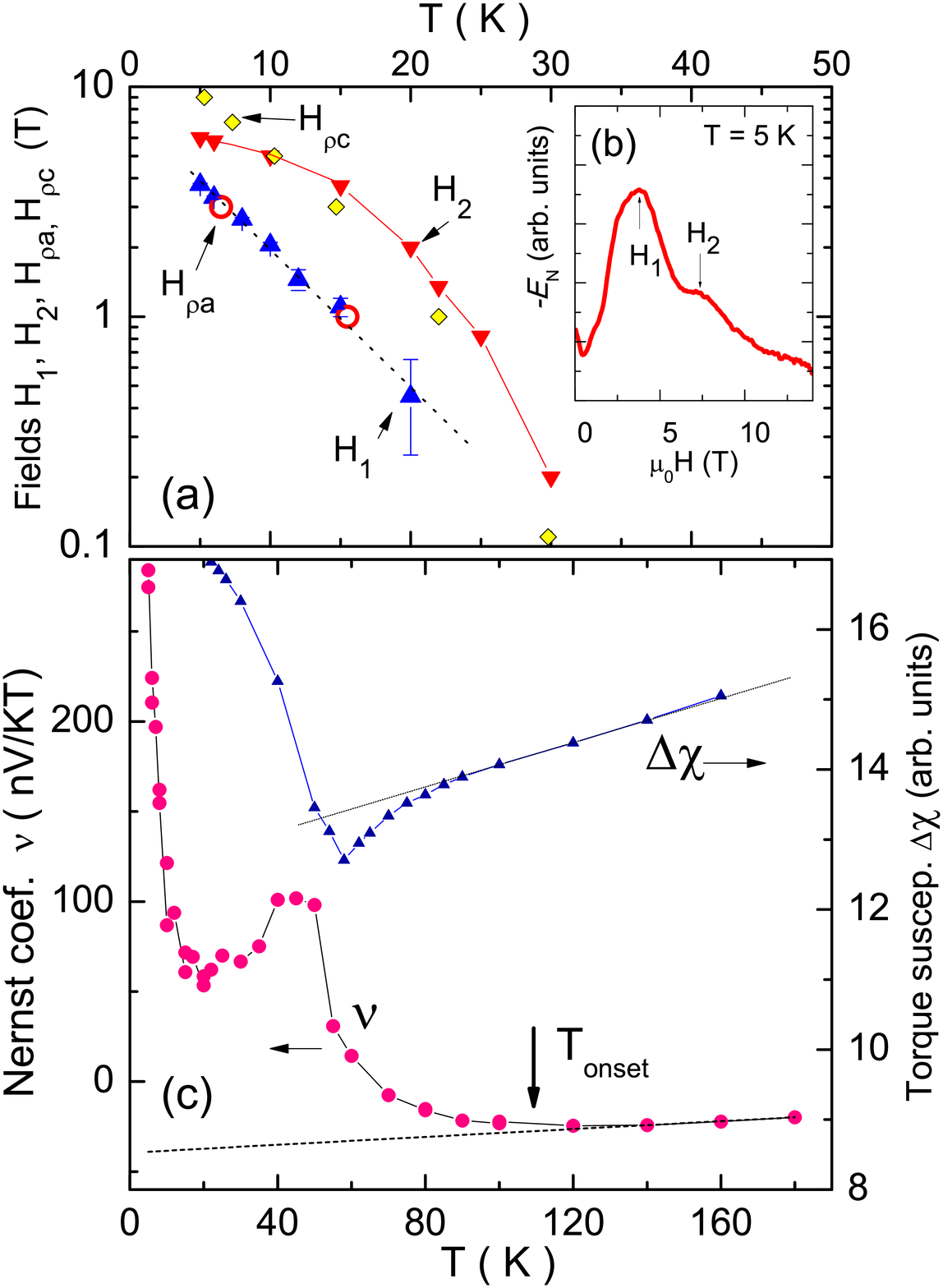}
\caption{\label{PeakH} (color online)
Panel (a): Semilog plot of the fields $H_1(T)$, $H_2(T)$, $H_{\rho b}(T)$ and $H_{\rho c}(T)$.
The data for $H_1$ (solid triangles) fits the form $H_0\exp(-T/T0)$ (straight line),
with $H_{0}$ = 8.28 T and $T_0$ = 6.86 K. $H_{\rho a}$ (open circles), inferred from 
$d\rho_{ab}/dT$~\cite{QLi07}, falls on the same line as $H_1$. $H_{\rho c}$ (open diamonds) obtained from $d\rho_{c}/dT$~\cite{QLi07} is roughly of the same field scale as $H_2$.
Panel (b) defines $H_1$ and $H_2$ for the curve $e_{N,even}$ at 5 K. 
Panel (c) displays the $T$ dependence of the Nernst coefficient $\nu = e^{N,odd}/H$ (solid circles)
and the torque susceptibility $\Delta\chi= \chi_c-\chi_a$ (solid triangles) measured 
with $\bf H$ (7 T) at 15$^{\rm o}$ to $\bf c$.
The increase in $\nu$ below $T_{onset}\sim$ 110 K correlates with
a diamagnetic contribution to $\chi$ from orbital currents. Below $T_{co}$,
the increase in $\nu$ is abruptly interrupted, but it resumes its steep increase below 20 K.
Below $T_{co}$, the large spin susceptibility obscures supercurrent
contributions to $\chi_c$.}
\end{figure}

Subtracting $kS(H,T)$ from $e^{obs}_{y,even}(H)$ at each temperature, 
we isolate $e_{N,even}(H,T)$, the field-even part of the 
intrinsic Nernst signal in Eq. \ref{eq:eN}.
The curves of $|e_{N,even}(H,T)|$ display broad peaks that shift to 
higher $H$ as $T$ decreases (Fig. \ref{EyHSH}b). The field at which the largest peak 
occurs is labelled $H_1(T)$. A smaller shoulder at higher field is labelled $H_2(T)$.
At 20 K, the weight in $e_{N,even}(H,T)$ is concentrated in a narrow trench
($|H_1|\sim$0.5 T). As $T$ is lowered, the two field scales $H_1$ and $H_2$ increase rapidly.
They correlate with distinct features in the in-plane resistivity $\rho_{ab}$ 
and the $c$-axis resistivity $\rho_c$. 
Below 40 K, the derivatives $d\rho_{ab}/dT$ 
and $d\rho_c/dT$ show maxima at the fields $H_{\rho a}(T)$ and $H_{\rho c}(T)$, respectively~\cite{QLi07}. 
In Fig. \ref{PeakH}a, we compare the $T$ dependences of $H_1$ and $H_2$ (solid symbols)
with $H_{\rho a}(T)$ and $H_{\rho c}(T)$ (open symbols)
(Panel (b) shows how $H_1$ and $H_2$ are defined). 
As shown, $H_1$ equals $H_{\rho a}$ within the resolution, while
$H_2$ is roughly of the same scale as $H_{\rho c}$.
Interestingly, $H_1(T)$ follows the Debye-Waller (DW) form $H_1 = H_0\exp(-T/T_0)$, with $T_0\sim$6.9 K. 
The DW form implies that thermally induced changes to the vortex system 
lead to prominent features in the anomalous Nernst signal $e_N(0,T)$.
In underdoped La$_{2-x}$Sr$_x$CuO$_4$, the DW form describes the 
melting field of the vortex solid (with comparable $T_0$)~\cite{LiNatPhys07}.  
We also note that the curves of $S$ vs. $H$ (dashed curves in Fig. \ref{EyHSH}a) 
display step-like increases when $H$ exceeds $H_1\sim H_{\rho,a}$, that match the abrupt increase 
in $\rho_{a}$. This pattern suggests that the collapse of the anomalous
Nernst signal at $H_1$ leads to an increase in dissipation and entropy flow. We return to this point below.

We field-antisymmetrize the Nernst curves in Fig. \ref{EyH} to obtain 
the conventional Nernst signal $e_{N,odd}(H) = \frac12[e^{obs}_y(H) - e^{obs}_y(-H)]$. 
The Nernst coefficient, $\nu = e_{N,odd}/H$ ($H$$\to$0), provides a useful comparison
between field-induced vortices and the spontaneous vortices. At high $T$ (120-180 K), $\nu$
is negative, reflecting the quasiparticle contribution to the Nernst signal 
(dashed line in Fig. \ref{PeakH}b). At the onset 
temperature $T_{onset}\sim$ 110 K, $\nu$ deviates from the dashed line and 
increases rapidly, as observed in La$_{2-x}$Sr$_x$CuO$_4$ (LSCO)~\cite{WangPRB06}. The deviation
correlates with an unusual downward deviation in the torque susceptibility $\Delta\chi=\chi_c-\chi_a$ 
in the torque signal (solid triangles), where $\chi_c$ ($\chi_a$) is the susceptibility
with $\bf H||c$ ($\bf ||a$). Above $T_{co}$, $\chi_c$ is $\sim 10\chi_a$~\cite{Hucker08},
so $\Delta\chi$ is dominated by $\chi_c$. Hence the downward deviation confirms the onset of
diamagnetic susceptibility in $\chi_c$ reported in Ref. \cite{Hucker08}. (Below $T_{co}$,
$\Delta\chi$ is complicated by a large local moment response in both $\chi_c$ and $\chi_a$.)
The magnetization results verify that, for $T>T_{co}$, the increase in $\nu$ arises from 
vortex fluctuations (and not from quasiparticles, as conjectured~\cite{LT}). 
A similar agreement between Nernst and torque experiments
was obtained for LSCO~\cite{WangPRB06,LiPRB09}.
At $T_{co}$, the increase in $\nu$ is abruptly interrupted. Below 20 K, 
however, $\nu$ resumes its steep increase as the condensate establishes 
long-range phase coherence.

The conventional field-antisymmetric Nernst signal 
shown in Fig. \ref{PeakH}b is generated by vortices introduced by an external $H$.
By contrast, we associate $e_{N,even}$ with vortices that are present in 
equilibrium at $H$ = 0, as in a 2D superconductor above $T_{BKT}$. However, unlike the BKT problem
(in which the net vorticity is zero in $H$=0), here we must have predominantly
``up'' vortices to produce a finite $e_N(0,T)$.
Using torque magnetometry, we have measured the irreversibility field $H_{irr}$
in the same crystal. As shown in Fig. \ref{EyT_ST}b, $H_{irr}$ 
has a very different profile from $e_N(0,T)$; $H_{irr}\to 0$ 
near 20 K, where $|e_N(0,T)|$ attains a maximum. Thus $e_N(0,T)$ is not caused by
field-induced vortices trapped in a non-equilibrium state. 
(In the interval 5 $<T<$ 20 K, the pair condensate rigidity is strongly 
inhomogeneous. The vortex solid exists in isolated regions detectable by
magnetization hysteresis. These regions do not contribute to the observed $e_N$ or $S$.)

The results in Refs. ~\cite{QLi07,TranquadaPRB08} have shown that pronounced
superconducting fluctuations extend from $T_{co}$ down to 5 K.  
The extreme anisotropy of this response 
indicates that the Josephson coupling between adjacent layers is highly frustrated.  
To explain this frustration, it has been proposed that pair-density-wave (PDW) 
superconductivity develops along with the stripe order \cite{Berg07,Berg09,Himeda02}.
Because the stripe modulation direction is orthogonal 
between adjacent layers, Josephson coupling cancels out.  
The abrupt interruption of the increasing trend in $\nu$ at $T_{co}$ (Fig. 4b) 
is consistent with a sharp change in the character of the probed phase coherence.  
Below $T_{so}$ = 40 K, previous results~\cite{QLi07,TranquadaPRB08,Valla06,He09},  
imply that competition between the PDW and uniform $d$-wave superconductivity exists.  
Eventually, at $\sim5$~K, the latter dominates and true 3D long-range phase coherence
prevails.  The steep rise of $\nu$ below 20~K is consistent with the eventual development 
of uniform $d$-wave order.

In the PDW state, small fluctuations in the 
Josephson phase $\theta({\bf r})$ about the uniform-phase state 
can lead to a gain in free energy~\cite{Berg09}.
The present results suggest to us that, below $T_{co}$, 
the sample spontaneously nucleates an array of 2D vortices in $H$=0, which
can provide a large phase-slip of $2\pi$. 
Having all the vortices be of the same sign (which breaks TRI) entails a 
cost in the kinetic energy of the supercurrent.
However, because the local supercurrent is weak, the cost may be offset by a large gain 
in condensate energy provided by
significant reductions in the interlayer phase frustration. Because $\theta$
is strongly fluctuating, we expect the vortices to flow freely in a gradient $-\nabla T$ 
and to generate a spontaneous Nernst signal.

The anomalous Nernst signal $e_N(0,T)$ attains its largest
amplitude at 20 K close to $T_{BKT}$ (16 K). Below $T_{BKT}$, the small but finite 
$\rho_{ab}$ implies that phase rigidity extends in the $a$-$b$ plane over 
sizeable lengths at $H$=0~\cite{QLi07}. However, when $H$ exceeds $H_{\rho a}$,
the collapse of the rigidity produces an increase in $\rho_{ab}$. As mentioned, this coincides with
a steep increase in $S$ which measures entropy flow (Fig. \ref{EyHSH}a),
as well as the collapse of $e_{N,even}$ above $H_1$ (Fig. \ref{EyHSH}b).
This suggests to us that the spontaneous vortices, when present, help
to establish a phase-coherent state that has low dissipation and low entropy. 
At the larger field $H_{\rho c}$, the step increase in $\rho_c$ signals the loss of interlayer
coherence. This is also reflected in $e_{N,even}$ as $H_2$, but as a much weaker
feature.

Despite the spontaneous nature of the time-reversal violation, some
external influence must nudge the system into selecting one direction in a given experiment.
We tried to change the sign by warming the sample to 290 K and then cooling in 
a different superconducting magnet, but it 
remained the same. We also tried field-cooling in $H$ = 14 T from 290 K, 
and also swept the field between +14 and -14 T both above and below $T_{co}$ but
could not alter the sign. A. Kapitulnik has suggested to us that a weak
magnetic ordering may onset at 360 K. Field-cooling from above 
360 K may pre-select the sign; this is left for a future investigation.

Recently, we learned of polar-Kerr rotation TRI violating results 
in LBCO-$\frac18$~\cite{KerrLBCO}. 
The Kerr angle $\theta_K$ in $H$ = 0 is unresolved from zero above $T_{co}$,
but increases abruptly at $T_{co}$, reaching a sharp maximum at 41 K.

We acknowledge valuable discussions with S. A. Kivelson, E. Fradkin, E. Berg and A. Kapitulnik.
We are indebted to Kivelson for valuable insights, and Kapitulnik 
for sharing unpublished Kerr results and suggestions. The research
is supported at Princeton by the U.S. National Science Foundation (Grant DMR 0819860),
and at BNL by the Office of Basic Energy Sciences, U.S. Department of Energy (Contract 
No. DE-AC02-98CH10886).

\emph{$^*$Present address of LL: Department of Physics, Univ. Michigan, Ann Arbor, MI}

\end{document}